\documentclass[]{mandm}

\usepackage{array}
\usepackage[utf8]{inputenc}

\usepackage{graphicx}
\usepackage{hyperref}
\usepackage{adjustbox}
\usepackage{subcaption}
\usepackage{float}
\usepackage{placeins}
\usepackage{stfloats}

\usepackage{bm}
\graphicspath{{./figures/}{./}}
\usepackage{upgreek}
\usepackage{amsmath}

\usepackage{balance}
\usepackage{todonotes}

\jourvolume{XX}
\jourissue{Y}
\jourpubyear{20ZZ}

\begin{document}

\sloppy

\title[Parallel mode differential phase contrast II]{Parallel mode differential phase contrast in transmission electron microscopy, II: K$_2$CuF$_4$ phase transition}

\author[G. W. Paterson et al.]{G. W. Paterson,$^1$
  G. M. Macauley,$^{1,2}$,
  S. McVitie,$^1$
  and Y. Togawa$^{3,1}$}

\affiliation{%
$^1$SUPA, School of Physics and Astronomy, University of Glasgow, Glasgow G12 8QQ, UK.\\
$^2$Current addresses: Laboratory for Mesoscopic Systems, Department of Materials, ETH Zurich, 8093 Zurich, Switzerland; Laboratory for Multiscale Materials Experiments, Paul Scherrer Institute, 5232 Villigen, Switzerland.\\
$^3$ Department of Physics and Electronics, Osaka Prefecture University, Sakai, Osaka, 599-8531, Japan.\\
Corresponding Author: G. W. Paterson \email{Dr.Gary.Paterson@gmail.com}}
  
\begin{frontmatter}

\maketitle

\begin{abstract}
In Part~I of this diptych, we outlined the theory and an analysis methodology for quantitative phase recovery from real-space distortions of Fresnel images acquired in the parallel mode of transmission electron microscopy (TEM).
In that work, the properties of the method, termed TEM-differential phase contrast (TEM-DPC), were highlighted through the use of simulated data.
In this work, we explore the use of the TEM-DPC technique with experimental cryo-TEM images of a thin lamella of a low temperature two-dimensional (2-D) ferromagnetic material, K$_2$CuF$_4$, to perform two tasks.
First, using images recorded below the ordering temperature, we compare the TEM-DPC method to the transport of intensity one for phase recovery, and discuss the relative advantages the former has for experimental data.
Second, by tracking the induction of the sample as it is driven through a phase transition by heating, we extract estimates for the critical temperature and critical exponent of the order parameter.
The value of the latter is consistent with the 2-D XY class, raising the prospect that a Kosterlitz--Thoules transition may have occurred.

\end{abstract}

\end{frontmatter}

\section{Introduction}
\label{sec:intro}
In Part~I of this two-part series~\citep{tem_dpc_part1}, we showed that the real-space distortions of Fresnel images taken in a transmission electron microscope (TEM) of samples supporting electro- or magneto-static fields may contain information sufficient for quantitative recovery of the electron-optical phase imparted on the imaging electrons.
The technique, which we refer to as TEM--differential phase contrast (TEM-DPC), has origins dating back decades~\citep{Fuller1960_jap_deflection}, but has gone largely unused since then.
In part, this is because previous approaches have used patterned films or shadow masks to form texture suitable for assessing the image distortions, which required modification of the microscope and limited mapping to stray magnetic~\citep{Wade_1976_ToM, Suzuki_1997, Suzuki_ToM_2000, Shimakura_JMSJ_2003} and electric fields~\citep{Sasaki2010_JEM_apertures} beyond the sample.
Another reason is the tremendous success since its early adoption in electron microscopy~\citep{Bajt_2000_TIE, DeGraef_JAP_2001_quant_TIE} of methods for phase recovery through solving the transport-of-intensity equation (TIE)~\citep{Teague_1983_greens_phase}.
This is a result of the TIE technique allowing phase characterisation across a sample and in vacuum in a standard TEM using data obtained at different defocus levels.

The fundamental principle behind the TIE and TEM-DPC methods is the same, but whereas TIE uses changes in intensity as the beam propagates, TEM-DPC uses changes in displacement as the beam propagates.
This difference endows each technique with distinct properties, and these are encoded in the principle equations.
For the transport of intensity approach~\citep{Paganin_PRL_1998_noninter_phase}, the TIE equation relates the phase, $\phi$, imparted by the sample on the electron beam of wavelength $\lambda$ to the rate of change in image intensity, $I$, with respect to defocus, $f$:
\begin{equation}
  \phi = - \frac{2\pi}{\lambda} \frac{1}{I_o} \nabla_\perp^{-2} \frac{\partial I}{\partial z}.
  \label{eqn:tie_simple}
\end{equation}
where the $z$-axis is taken to lie anti-parallel to the beam propagation direction, $I_o$ is the in-focus image intensity, $\nabla_\perp^{-2}$ is the inverse Laplacian, and the subscript indicates the plane perpendicular to the beam.
This equation is a simplified form, valid in cases where the in-focus image intensity is spatially uniform (we use the full equation later in our data analysis).
For TEM-DPC, the phase gradient is related to the lateral deflection, $\bm{\Delta r}_\perp$, at defocus $\Delta f$ by:
\begin{equation}
  \nabla_\perp \phi = -\frac{2\pi}{\lambda} \frac{\bm{\Delta r}_\perp}{\Delta f},
  \label{eqn:phasegrad_displacement}
\end{equation}
where $\nabla_{\perp}$ is the gradient operator.
One of the most notable differences in these equations is the absence of image intensity terms in the TEM-DPC one.
As a result of this, the TEM-DPC technique is intrinsically less sensitive to systematic errors arising from changes in illumination than is the TIE method.
There are intensity factors that influence the ability to extract $\bm{\Delta r}_\perp$, and we refer to the accompanying part of this work for a detailed discussion of them~\citep{tem_dpc_part1}.

We discussed different methods for quantifying $\bm{\Delta r}_\perp$ and reported a methodology based on well-established non-rigid image alignment methods~\citep{Kroon2009_demon_reg, Jones2015_smart_align} in Part~I.
This approach makes use of structural contrast originating from small intrinsic imperfections that are present in almost all samples and, consequently, enables phase extraction across an entire sample without modification of the microscope.
Using simulated data from different magnetic samples, we showed that the TEM-DPC technique has further potential benefits over the TIE one in some cases.
One such case arises when the structural contrast is particularly large compared with that from magneto- or electro-static fields supported by the sample.

In this work, we apply the TEM-DPC technique and methodology developed in Part~I to experimental cryo-TEM Fresnel images from a sample of K$_2$CuF$_4$ that contains small but important imperfections naturally absent from idealised simulations.
This material may be regarded as one of several layered Heisenberg systems with planar anisotropy~\citep{Bramwell_1993_2D_XY_beta} which behave as quasi 2-D spin systems within certain temperature windows.
For K$_2$CuF$_4$, this window is at low temperatures (the exact window is discussed in detail later), where a modified Kosterlitz--Thouless (KT) transition was observed through neutron scattering and static magnetisation measurements reported almost 40 years ago~\citep{Hirakawa_JAP_1982}.
In our recent work on this material~\citep{Togawa2021_JPSJ_transition}, we reported real-space, direct observations of vortex and anti-vortex structures at a single temperature, consistent with the excitation types expected from the KT transition~\citep{Berezinskii_JETP_1971_2D_1, Berezinskii_JETP_1972_2D_2, Kosterlitz_1973}, but possibly modified by anisotropies arising from the finite sample size.

In the following, we briefly summarise the relevant measurement details before comparing the TIE and \mbox{TEM-DPC} methods using data from below the ordering temperature.
We find that the latter technique can have advantages at sample edges, and also when additional image distortions from imperfection in the microscope optics are present.
We then apply the TEM-DPC method to the analysis of data acquired at multiple temperatures across the phase transition in order to characterise the magnetic textures during the transition and extract estimates for the critical temperature and critical exponent of the ordering parameter.
The values of the critical parameters are consistent with previous reports~\citep{Hirakawa_1973_JPSJ} and are suggestive of a material with a \mbox{2-D} XY character.
These results confirm that the TEM-DPC technique and analysis methodologies presented constitute a valuable tool with prospects for use in the characterisation of wide range of samples in materials science.

\section{Experiment}
The sample under study was taken from a bulk crystal of K$_2$CuF$_4$ and thinned to an electron-transparent thickness of $\sim$150~nm using a standard focused ion beam preparation technique~\citep{Schaffer2012_um_fib_sample_prep}.
Imaging of the liquid He cooled sample was performed in an Hitachi \mbox{H-1000FT} electron microscope equipped with a highly coherent cold-field electron source operated at an acceleration voltage of 1~MV~\citep{Kawasaki_2000_APL_1MeV_TEM}.
Multiple datasets were acquired at fixed and variable temperatures by recording videos with a Gatan Orius camera at 1~frame/sec using Gatan Digital Micrograph\textsuperscript{TM} for later analysis.
With an out-of-plane field of 100~Oe applied to the sample, it was found to be uniformly magnetised.
In the following, we focus on a dataset where the sample was driven through a phase transition by heating the sample from 5~K to 7~K, with an out-of-plane field of 25~Oe applied to the sample (pointing into the page in the images shown later), and a defocus of 44~mm.
In this dataset, the pixel size was 10.46~nm, which is much smaller than needed to resolve the magnetic texture in the Fresnel images, but is suitable to resolve details of the non-magnetic contribution to the image contrast.

\section{CTEM and Fresnel Images}

\begin{figure*}[!tbh]
  \centering
      \includegraphics[width=15cm]{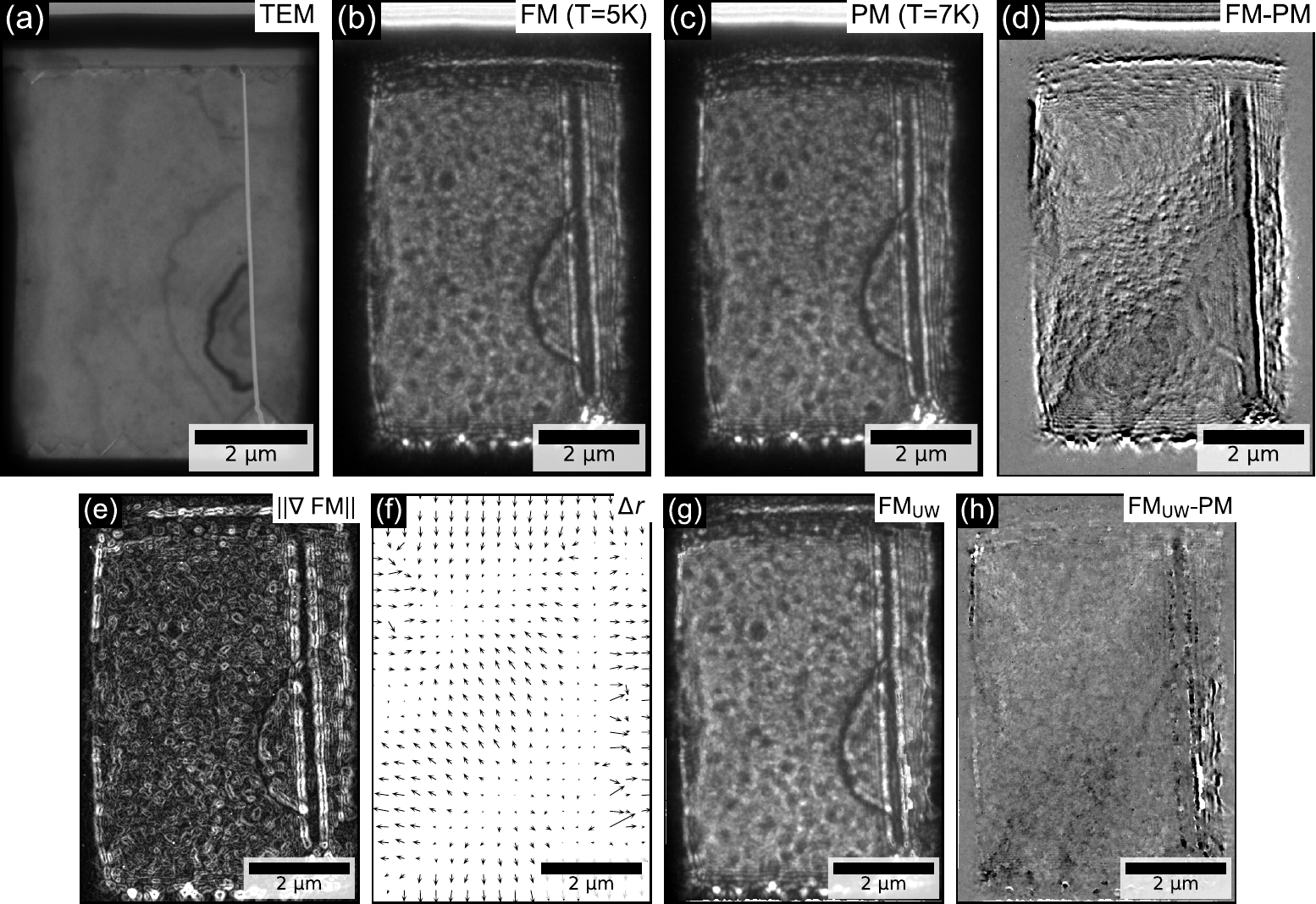}
      \caption{Cryo-TEM images of a K$_2$CuF$_4$ sample.
      In-focus (a) and underfocus (44~mm) Fresnel (b, c) images at temperatures (b) below and (c) above the phase transition where the sample is ferromagnetic (FM) and paramagnetic (PM), respectively (versions of (b) and (c) are also included in \citep{Togawa2021_JPSJ_transition}).
      (d) Difference Fresnel image, showing a combination of high-frequency non-magnetic texture and magnetic induction induced intensity variations suggestive of the presence of a clockwise vortex at the bottom and an anticlockwise vortex at the top of the sample.
      (e) The derivative magnitude for the FM image in (b), which amplifies the non-magnetic contrast during the distortion analysis.
      (f) The distortion field that maps (c) to (b), obtained with a regularisation size of 2~pixels.
      The maximum magnitude is $\lesssim$5~pixels; the image size is 528$\times$774 pixels; and only 1 in 30 arrows are shown along each axis.
      (g) The FM image of (b) with the magnetism induced image distortion removed using the distortion field in (f).
      The local image intensities are maintained in this procedure.
      (h) The equivalent difference Fresnel image of (d) made with the unwarped image of (g), showing clear magnetic intensity variations due to cancellation of the non-magnetic contrast in the aligned images.
      The plotted intensity scale in (d) and (h) are the same.}
      \label{fig:TEM_images}
\end{figure*}

The top row of Figure~\ref{fig:TEM_images} shows conventional TEM (CTEM) and Fresnel images of the K$_2$CuF$_4$ sample under study.
In the in-focus image [Fig.~\ref{fig:TEM_images}(a)], the bright straight vertical line towards the right hand side of the sample and the two jagged bright lines at the top and bottom of the sample are cracks that occurred during temperature cycling.
These conveniently allow the film to remain largely strain free and also break the exchange coupling on all three edges, defining the area of the sample on the left hand side that we will investigate.
The left hand edge of the transparent section is connected to a much thicker region of the material, allowing good thermal contact to the cooled sample holder.
The semi-circular dark ring segments are diffraction effects in the single crystal material as a result of bend contours.
The width of the sample was determined from this image to be 5~$\muup$m, and this value was used as a reference to calibrate the pixel size of all defocused images.

Fresnel images at temperatures below [Fig.~\ref{fig:TEM_images}(b), T = 5~K] and above [Fig.~\ref{fig:TEM_images}(c), T = 7~K] the phase transition show significant structural contrast.
The defocus value of 44~mm was determined from the contrast transfer function minima in reciprocal space using an automatic procedure discussed in the appendix and depicted in Figure~\ref{fig:defocus}.
This defocus value is an order of magnitude larger than that typically used for permalloy thin films in a 200~kV microscope, and was used to provide suitable image intensity variations and distortions for the smaller wavelength electrons at the higher acceleration voltages used here ($\left\lVert \bm{\Delta r}_\perp \right\rVert \propto \lambda \Delta f$, from Eq.~(\ref{eqn:phasegrad_displacement})).
At such large defocus values, changes in magnification and barrel distortion are common artifacts that can arise from the microscope optics.
These Fresnel images were recorded at the same nominal magnification as the in-focus image and the same number of pixels are shown for all images in the top row of Fig.~\ref{fig:TEM_images}.
The smaller pixel size of the sample and the rounded edges at defocus make it clear that these artifacts are present in the experimental Fresnel data.
Another factor which will influence our analysis is the change in the apparent position of the bend contour with changes in focus [c.f. Figs.~\ref{fig:TEM_images}(a) and \ref{fig:TEM_images}(b)].
This occurs as a result of small changes in the illumination conditions and, as we will see later, most strongly affects the TIE analysis.

The most striking feature of the Fresnel images is the strong non-magnetic contrast arising from fine surface texture.
Only the barest hints of this are visible in the CTEM image [Fig.~\ref{fig:TEM_images}(a)], suggesting its origin is a mixture of phase and amplitude modulations.
However, also present but barely perceptible by eye in the Fresnel images is a very weak intensity contrast from the magnetic induction of the sample.
Some details of the magnetic contrast may be seen in the difference image [Fig.~\ref{fig:TEM_images}(d)].
Before this image could be calculated or the data analysed, the image stack was aligned rigidly using each entire image, including the support frame, in order to remove lateral sample drift during heating.
Were the sample intact, standard Fourier approaches to image registration could have been used.
However, because different parts of the cracked sample move at different rates, we instead used the scale-invariant feature transform (SIFT)~\citep{Lowe2004_sift} provided in the Fiji image processing package~\citep{fiji_2012} to rigidly align our data.
This approach matches keypoints between images and, in our case, preferentially aligned the K$_2$CuF$_4$ film.
In cases where the sample moves as one and the support structure is transparent or has well defined edges, this approach can also be of use by allowing only the fixed support structure to determine the alignment, thereby avoiding any bias from regions of the images undergoing distortion due to the sample induction.

In the difference image [Fig.~\ref{fig:TEM_images}(d)], the presence and circulation direction of vortices may be inferred from the high frequency image texture and the relative brightening and darkening of the vortex cores, respectively.
The locations of the vortices are marked by regions of small high-frequency signal amplitude and azimuthal lines of contrast.
At those sites, the image deformation is low because a component of the moment lies out of plane and, thus, deflection of the beam is reduced.
The darkening (brightening) at the bottom (top) of the image indicates that the vortices are clockwise (anticlockwise) as a result of divergence (convergence) of the beam.
The changes in intensity from the magnetic contrast is $\sim$7~\% of the average intensity, while the structural intensity range is $\sim$$10\times$ larger than this.
Even though the two signal strengths are suitable for the TIE and TEM-DPC analyses, the much larger structural contrast potentially allows for much faster acquisitions with lower doses, which may be advantageous for time resolved studies.

The bottom row of Fig.~\ref{fig:TEM_images} shows the results of calculations performed on the top row data, with a small crop to remove unused data.
To enhance sensitivity to the high frequency structural contrast and minimise the influence of the low frequency magnetic contrast in the \mbox{TEM-DPC} analysis, image derivatives were used as the source and target images between which the distortion fields were calculated; an example derivative image is shown in Fig.~\ref{fig:TEM_images}(e) (corresponding to the FM data in Fig.~\ref{fig:TEM_images}(b)).
We note that the non-rigid analysis methodology developed in Part~I of this work and used here allowed the inclusion of the entire sample and support structure areas in these analyses.
This was not possible in our earlier work~\citep{Togawa2021_JPSJ_transition} on account of the choice of \mbox{B-spline} basis~\citep{bunwarpj_2006, fiji_2012} for the distortion field.

The image distortion field extracted from the FM and PM images after differentiation [Fig.~\ref{fig:TEM_images}(f)] clearly shows the vertical crack and the regions where the two magnetic vortices of opposite circulation direction are acting as converging or diverging lenses at the top and bottom of the sample, respectively.
The maximum vector field magnitude is relatively small at $\lesssim$5~pixels, approximately equal to 1\% of the 528~pixel wide image or, equivalently, to a real space distance of $\sim$50~nm.
As a result, the FM image after applying the inverse of the mapped deformation field [Fig.~\ref{fig:TEM_images}(g)] appears very similar to the original image [Fig.~\ref{fig:TEM_images}(b)] by eye.
However, as the `unwarping' process used was interpolation of the image at the new coordinates, the local image intensity is maintained while the shape is subtly changed.
The difference [Fig.~\ref{fig:TEM_images}(h)] between this unwarped FM image and the PM reference image  shows that almost all high frequency non-structural components are now absent, allowing the weak intensity changes between images from the magnetic vortices to be more clearly observed.

The image distortion field in Fig.~\ref{fig:TEM_images}(f) and in all experimental TEM-DPC analysis was determined using non-rigid alignment to a reference image from the sample in the PM regime, and with a Gaussian regularisation $\sigma$ of 2~pixels, corresponding to a full-width-half-maximum size of $\sim$49~nm.
Note that the distortion resolution is much better than this contribution to the spatial resolution, and that other factors will also influence the overall spatial resolution.
Further discussion and examples of the regularisation and its influence on the signal strength and noise level, and on the spatial resolution is given in the Appendices (see Figure~\ref{fig:TEM_align} and related discussion).
Next, we will compare the TEM-DPC and TIE methods to extract the sample induction at base temperature, before considering intermediate and higher temperatures.

\section{TIE and TEM-DPC Induction Maps}
In all following analyses, we extract the sample electron-optical phase using Eq.~(\ref{eqn:phasegrad_displacement}) defined in the introduction for the TEM-DPC method, and the full version of the TIE one for that method:
\begin{equation}
  \phi = - \frac{2\pi}{\lambda} \nabla_\perp^{-2} \left\{ \nabla_\perp \cdot \left[ \frac{1}{I_o} \nabla_\perp \nabla_\perp^{-2} \frac{\partial I}{\partial z} \right] \right\}.
  \label{eqn:tie}
\end{equation}
For both techniques, we convert the extracted phase (or phase gradient) to sample induction using the appropriate equations (defined in Part~1) for a more direct interpretation of the data.

\begin{figure*}[tbh]
  \centering
      \includegraphics[width=14.5cm]{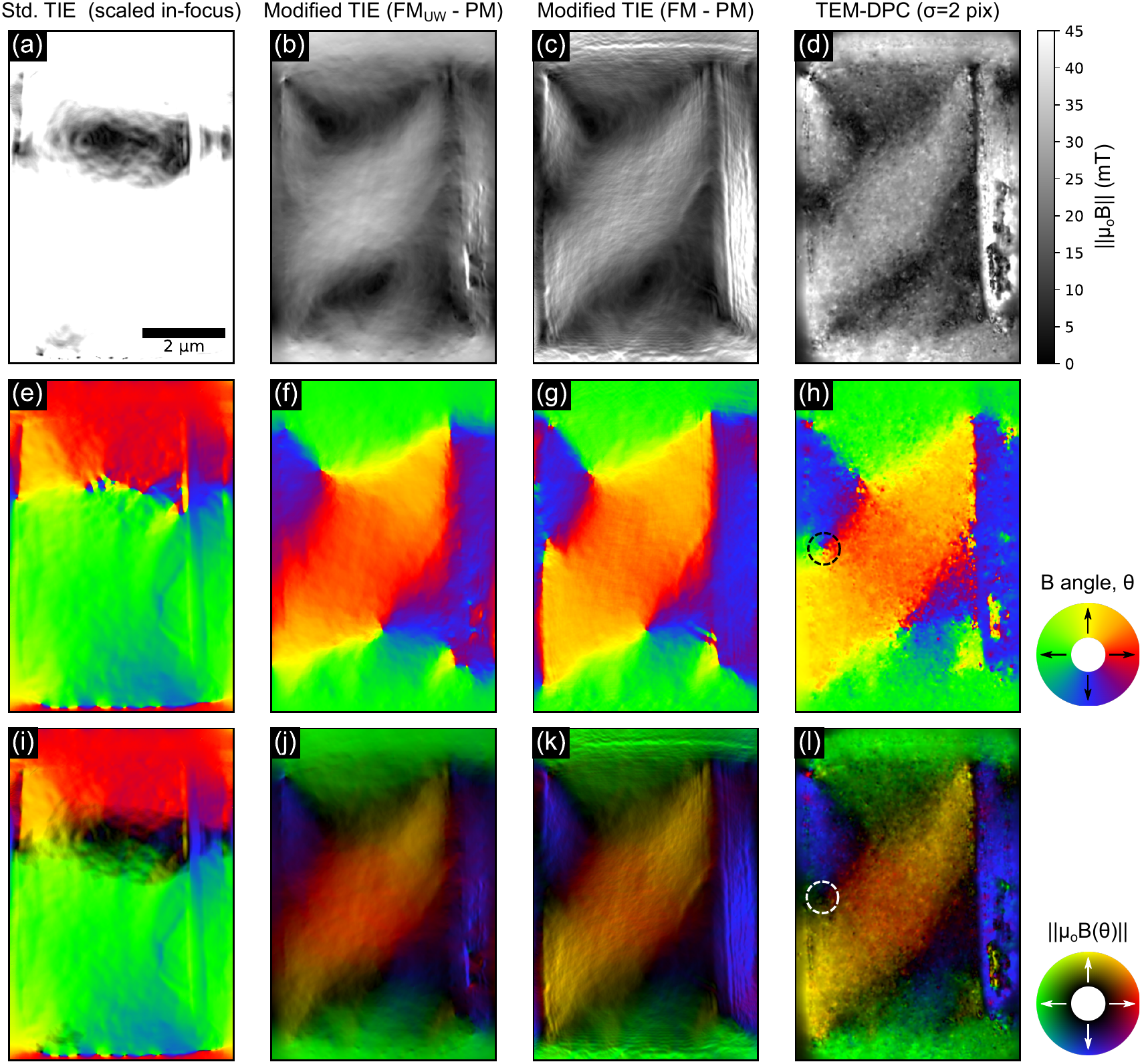}
      \caption{Comparison of induction maps for the K$_2$CuF$_4$ sample at 5~K, from different TIE methods and the TEM-DPC one applied to the data in Fig.~\ref{fig:TEM_images}.
      The different analysis methods are shown in columns, as indicated by the column titles.
      The rows show the vector field magnitude (top), vector field angle (middle), and a combination of the two (bottom).
      Each row is plotted on the same scale, indicated by the color bar or colour wheel against each row.
      The discontinuity in the TEM-DPC data to the right of the crack arose from a misalignment which went uncorrected.
      The dashed circles in the TEM-DPC data mark the location of an antivortex.}
      \label{fig:B_vector_plots}
\end{figure*}

Figure~\ref{fig:B_vector_plots} shows the induction maps from analysis of the data in Fig.~\ref{fig:TEM_images} plotted as magnitude (top row), angle (middle row), and a combination of the two (bottom row).
The direction of the induction is indicated by the hue, while the brightness (black to colour) indicates the vector magnitude, as shown by the colour wheels.
All panels have the same scaling from 0 (black) to 45~mT for full colour, calculated from the projected induction using the known sample thickness.
The first three columns show data from the TIE method using different input images, while the fourth column shows the results of the \mbox{TEM-DPC} analysis.

The first column of Fig.~\ref{fig:B_vector_plots} shows the results of the conventional TIE analysis, using the FM image and the in-focus image after scaling it to $\sim$0.9$\times$ of its original size in order to match the scale of the FM image.
The vector field magnitude [Fig.~\ref{fig:B_vector_plots}(a)] is almost completely saturated due to systematic errors.
The mean intensities of the Fresnel images used differ by only $\sim$0.5~\% and normalising the intensities makes little difference; this error is due to subtle changes in the distribution of intensity rather than a net change in brightness of the illumination and it thus also influences the details of the extracted induction.
In addition to this effect, the images show significant texture due to structural contrast, and multiple strong bend contour features are visible in the angle data [Fig.~\ref{fig:B_vector_plots}(b)], due to changes in their position and intensity between images.

A natural choice to improve the conventional TIE analysis is to replace the in-focus image with the PM one, so that the illumination changes are minimised, and to use the unwarped version of the FM image so that structural contrast cancels, as we saw in Fig.~\ref{fig:TEM_images}(g).
The problem with this approach is that structural contrast from using the PM image in place of $I_0$ in Eq.~(\ref{eqn:tie}) will still influence the results.
This can be overcome in cases where the structural contrast in the PM image can be removed by filtering.
In this process, it is important to maintain the overall intensity variations from the non-structural contrast, which will often be possible in thin film samples.
Our dataset has many thickness and intensity changes and is particularly difficult to filter, so to demonstrate this approach we simply use as $I_0$ a flat image with values taken from the 50 percentile value of the PM image.
As a result, we expect some details of the induction will not be accurately recovered, particularly at locations of changing intensity such as near the sample edges.
The results of doing this are shown in the second column of Fig.~\ref{fig:B_vector_plots}.
This is the first dataset to show the clear presence of two vortices, and it does indeed produce TIE results with the minimum influence of structural contrast.

The third column of Fig.~\ref{fig:B_vector_plots} shows TIE analysis results from a modified method using the PM and FM images with no unwarping.
These plots are more mottled than those in the second column, but have improved sharpness at the sample edges.
The third column data compare well with the TEM-DPC data in the fourth column, showing the validity of the TEM-DPC analysis methodology presented which, we emphasise, uses largely independent information from the TIE method: displacement rather than intensity.
This level of agreement was only possible because the microscope optics and electron source were relatively stable during the experiments; if they were not, then even the modified TIE methods would show systematic errors.

Perhaps the most obvious difference between data from the TEM-DPC method and that from the most accurate TIE one [third column of Fig.~\ref{fig:B_vector_plots}] is the noise level, where the data from the former has a slightly more speckled appearance due to very small imperfections in the alignment of the structural features.
All of the TIE data appear smooth due to the filtering effect of the Fourier space methods used to perform the analysis, which reduces the spatial resolution.
This effect, in combination with the image distortion, can be seen by comparing the magnitude and angle plots towards the top of the vertical crack; while the TIE data show smooth and continuous profiles, the TEM-DPC data is sharp and much more discontinuous, giving crisp edges.
A further improvement in the TEM-DPC method is the reduction in Fresnel fringes in the induction data from the rapidly varying sample thickness, which may be seen by comparing the bottom edges of Figs.~\ref{fig:B_vector_plots}(k) and \ref{fig:B_vector_plots}(l).
The main reason for this is that the influence of intensity contrast is suppressed in favour of structural contrast (see Fig.~\ref{fig:TEM_images}(e)), which is also the reason for the reduced influence of the bend contour in the TEM-DPC data (\textit{cf.} Figs.~\ref{fig:B_vector_plots}(g) and \ref{fig:B_vector_plots}(h)).

Looking now at the magnetisation topology, the vortices seem to be somewhat modified by the finite sample size, each approaching a Landau-like state with approximately 90$\mathrm{^o}$ walls in the corners.
The magnetisation mostly lies parallel to the edges formed by the cracks which break the exchange coupling (there may be some dipolar coupling across the vertical crack).
The resulting configuration resembles the diamond-state seen in rectangular shaped structures due to shape anisotropy.
However, looking in detail at the TEM-DPC data, we can see it is only a resemblance: the pattern is offset from the sample edges; at the center of the left hand edge of the transparent region is a complete anti-vortex [circled in Figs.~\ref{fig:B_vector_plots}(h) and \ref{fig:B_vector_plots}(l)]; and the magnetisation configuration continues into the thicker section of material to the left.
The relatively sharp angular transition marked by a blue to red/orange transition between this anti-vortex and the top vortex [Fig.~\ref{fig:B_vector_plots}(h)] suggests that these structures are connected, and it has been suggested that these may have formed through a KT transition as the sample was cooled~\citep{Togawa2021_JPSJ_transition}.
A repeat measurement with 10~Oe applied out-of-plane and a defocus of 19~mm showed similar vortex and anti-vortex structures to those shown here, but of the opposite chirality and located in slightly different positions.
Next, to further test for evidence of a KT-like transition, we examine data at multiple temperatures acquired under the same experimental conditions as that used for the data in Fig.~\ref{fig:B_vector_plots}.

\section{Phase Transition}
Having compared the different analysis methods above, we now explore details of the phase transition.
K$_2$CuF$_4$ has been shown to have multiple regimes of spin coupling at different temperatures and fields.
As the temperature is decreased at zero field, a transition from a 2-D Heisenberg to 2-D XY regime has been estimated to occur at 7.3~K, with 3-D correlations beginning to develop below 6.6~K, leading to a 3-D XY regime~\citep{Hirakawa_JAP_1982}.
In almost all materials, multiple effects may influence the results around the critical point(s), including crystalline and sample shape anisotropy, dipolar interactions, finite in-plane size effects, finite thickness, external fields, and domain configuration.
These effects are discussed in detail in a review by \cite{Vaz_RPP_2008_review}, and are likely to influence the properties of our sample to some degree, as discussed in our earlier work~\citep{Togawa2021_JPSJ_transition}.
While keeping this in mind, it is nevertheless worth extracting what information we can from the experimental data.

\begin{figure}
  \centering
      \includegraphics[width=8cm]{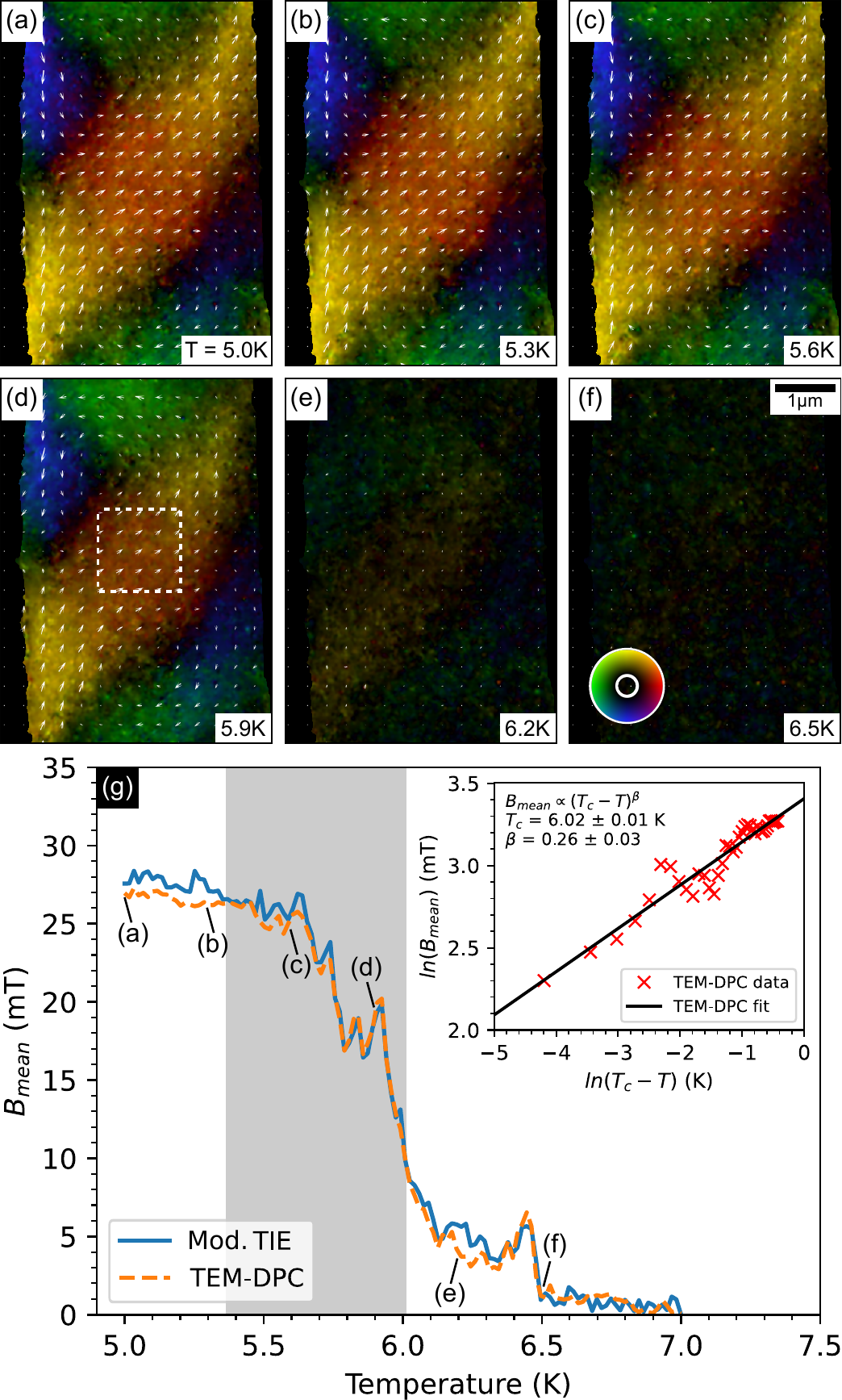}
      \caption{(a)-(f) Examples of the induction maps from the \mbox{TEM-DPC} method applied to data across the FM to PM phase transition at the temperatures indicated in the annotations.
      The induction vector field magnitude and directions are indicated by the brightness ([0, 45]~mT) and hue, respectively, of the colour wheel inset in (f).
      The critical temperature is approximately 6.0~K, as shown in (g).
      (g) Temperature dependence of the mean induction, $B_{mean}$, in the 130$\times$130 pixel region between the vortex cores marked by the white dashed box in (d) for the modified TIE method and the TEM-DPC method.
      The inset shows the logarithmic transform of the TEM-DPC data from the shaded region of the main panel (symbols) and a fit to it (black solid line) of Eq.~(\ref{eqn:exponent}) to extract estimates of the critical temperature, $T_c$, and critical exponent, $\beta$.
      The errors shown are from the quality of the fit, to which a calibration error of 0.1~K is added in the main text.}
      \label{fig:temperature_dependence}
\end{figure}

Figures~\ref{fig:temperature_dependence}(a)-\ref{fig:temperature_dependence}(f) show the TEM-DPC induction maps at different temperatures across the phase transition, with the data either masked beyond the sample edges or cropped near their locations.
A video of the full transition is included in Supplemental, and is further described in the Appendices.
As the sample is heated through the phase transition, the magnetisation and thus induction reduces in magnitude and the vortex structures break down as the sample becomes paramagnetic (strictly speaking, there may remain a small amount of magnetisation which is stabilised by the out-of-plane external field, to which our measurements are insensitive).
To track the transition, we use as a proxy for the order parameter the mean induction, $B_{mean}$, in the region between vortex cores marked by the white dashed box in Fig.~\ref{fig:temperature_dependence}(d), where the magnetisation is approximately divergence-free.
This quantity is plotted as a function of temperature, $T$, in Fig.~\ref{fig:temperature_dependence}(g) for both the TIE method using the FM and PM images directly, and the TEM-DPC method, which agree very well with one another, with both traces showing a clear transition at a temperature of around 6~K.
A separate measurement with a field of 50~Oe applied to the sample at a fixed temperature of 6.1~K showed variations in contrast with time, indicative of thermally activated oscillations between different states.

Earlier neutron diffraction measurements of the temperature dependence of the magnetisation of K$_2$CuF$_4$ found that it followed Bloch's $T^{3/2}$ law at low temperatures, with a saturation magnetisation $B_s = \mu_0 M(T=0)$ = 123.4~mT and a Curie temperature, $T_c$, of 6.25~K~\citep{Hirakawa_1973_JPSJ}.
Our measurements do not reach low enough temperatures to accurately estimate $B_s$ from the data.
In our case, the maximum value plotted of $\sim$30~mT, may also be influenced by a combination of the broad nature of the vortex cores and their proximity to one another, both of which will reduce the measured induction at a given temperature.
The finite thickness of the sample may also modify the magnetic configuration, as discussed in detail in our earlier work~\citep{Togawa2021_JPSJ_transition}.

Driven through a phase transition, the relaxation time of a system diverges and the properties of the out-of-equilibrium system are characterised by intrinsic interactions, and fall within one of several universality classes.
As a real, finite system is enlarged, the properties will approach those of the class to which it belongs, and are uniquely described by a set of critical exponents.
At temperatures close to the phase transition, the magnetisation and thus $B$ should follow~\citep{Vaz_RPP_2008_review}: 
\begin{equation}
    B(T) \propto \left( T_c - T \right)^\beta,
    \label{eqn:exponent}
\end{equation}
where $\beta$ is a critical exponent for the ordering parameter, and $T_c$ is a critical temperature.
It has long been known that for an isotropic 2-D XY-system, the system must be finite for ordering to occur at a finite temperature~\citep{Mermin_Wagner_1966}.
However, consistent with theory, many experimental quasi 2-D XY magnetic systems have been found to show universal scaling with a $\beta$ value of 0.23~\citep{Bramwell_1993_2D_XY_beta}.
Although a range of $\beta$ values are reported from experiments on different materials, in part as a result of their differing crystal field strengths~\citep{Taroni_2008_crit_exp_window}, the value in K$_2$CuF$_4$ has been measured to be 0.22 at temperatures close to $T_c$~\citep{Hirakawa_1973_JPSJ}, consistent with the 2-D XY class.
At lower temperatures, the value has been found to increase to 0.33$\pm$0.03~\citep{Hirakawa_1973_JPSJ}, close to the $\beta$ values of classes with a 3-D character (3-D Heisenberg: 0.369~\citep{Campostrini_prb_2002_heisenberg_crit}; 3-D XY: 0.349~\citep{Campostrini_2001_PRB_3DXY_crit_exp}).

While a full study of all factors that may influence the properties extracted from our measurements of this system would be extremely challenging and lies beyond the scope of this work, we may gain some insight into how $\beta$ and $T_c$ are affected by the magnetisation topology by fitting Eq.~(\ref{eqn:exponent}) across the dataset to the net induction in 10$\times$10~pixel subsets, in each of which the induction is relatively uniform.
Away from the sample edges and the vortex cores, the values and errors are relatively uniform and it is from here, in the centre between the two vortices, that the induction profiles of Fig.~\ref{fig:temperature_dependence}(g) were taken.
As the magnetisation in this region lies in a relatively uniform direction, the result of analysing the average of all pixels from this area produces very similar results to the average of the extracted spatially resolved parameters.
The results of fitting Eq.~(\ref{eqn:exponent}) to the averaged data is shown by the solid line in the inset to Fig.~\ref{fig:temperature_dependence}(g), and gives values of $T_c = 6.0 \pm 0.1$~K and $\beta = 0.26 \pm 0.03$, where the error in the former is from the temperature calibration, while that in the latter is from the quality of the fit.
The temperature range used in this fit was determined by how well the data conformed to Eq.~(\ref{eqn:exponent}) and is shown by the grey background in Fig.~\ref{fig:temperature_dependence}(g).
Our estimate of $T_c$ agrees well with the previously reported value.
Although the error in the extracted $\beta$ value is higher than would be ideal, the value itself lies closer to that expected of the 2-D XY class than it is to those of 3-D systems, and thus is suggestive of a system with 2-D XY character.

Finally, it is worth commenting briefly on the direct observations of the phase transition.
In the pure \mbox{XY-class}, the transition at finite temperatures is described by the KT mechanism~\citep{Kosterlitz_1973}, where bound vortex and anti-vortex pairs with zero net vorticity form and mediate the transition.
These bound pairs may be mobile below the phase transition temperature and, when driven towards higher temperature, are expected to unbind at the KT temperature~\citep{Pokrovsky_JMMM_1999_2d}, while thermal spin fluctuations grow.
Ideally then, to unambiguously observe a KT transition in real-space alone, the sample should be large compared to the scale of the defects that form during the transition, which is unfortunately not the case here.
While this does not exclude the possibility that the vortices and anti-vortices formed during a KT transition (indeed, the exponent $\beta$ is suggestive of 2-D XY character), further work is required to fully understand all aspects of the observed behaviour.
Ideally, this work would be done with thinner samples of a greater lateral extent which would allow direct visualisation of more dynamics during the phase transition, and thus access to other properties and critical exponents, the correlation function, in particular.

\section{Conclusions}
Using the TEM-DPC technique for quantitative phase recovery from Fresnel images and the non-rigid image alignment tools reported in Part~I of this work~\citep{tem_dpc_part1}, we have characterised the induction of a thin lamella of K$_2$CuF$_4$, a material of increasing interest for the quasi 2-D ferromagnetism it supports.
We have compared the TEM-DPC technique with the TIE one using the same data obtained below the K$_2$CuF$_4$ ordering temperature, and shown that the former method has potential advantages over the latter in the analysis of experimental data.
This includes in areas near to sample edges, where field-induced distortion of the images breaks the assumptions of TIE method, and also in the common situation where small unavoidable changes in illumination or image distortion arise from the microscope optics.

By mapping the K$_2$CuF$_4$ sample induction through a temperature-driven phase transition with the TEM-DPC method, we have extracted estimates for the critical temperature, $T_c$, and critical exponent, $\beta$, of $6.0 \pm 0.1$~K and $0.26 \pm 0.03$, respectively.
These values agree with earlier reports using different techniques and are suggestive of the sample having a 2-D XY character.
This, together with the direct observation of vortex and anti-vortex structures, gives further support to the suggestion that a \mbox{KT-like} transition~\cite{Togawa2021_JPSJ_transition} has occurred.

This initial experimental demonstration of the \mbox{TEM-DPC} technique shows that it works well to extract the features of a magnetic sample, and has the potential to characterise other material systems and sources of magnetic or electric fields through recovery of the electron-optical phase.

Original data files for the work reported herein are available at DOI: TBA.
All substantive simulation, analysis, and visualisation code is freely available in the open source \texttt{fpd} Python library~\citep{fpd}.

\noindent\small\color{Maroon}\textbf{Acknowledgements }\color{Black}

We are grateful to Prof. Kinshiro Hirakawa of the University of Tokyo for fruitful discussions and for providing us with K$_2$CuF$_4$ crystals.
We acknowledge support from Grants-in-Aid for Scientific Research on Innovative Areas `Quantum Liquid Crystals' (KAKENHI Grant No. JP19H05826) from JSPS of Japan, and Grants-in-Aid for Scientific Research (KAKENHI grant Nos. 17H02767 and 17H02923) from JSPS of Japan.
This work includes results obtained by using the research equipment shared in the MEXT Project for promoting public utilisation of advanced research infrastructure (Program for supporting introduction of the new sharing system: Grant Number JPMXS0410500020).
We acknowledge support from the Engineering and Physical Sciences Research Council (EPSRC) of the United Kingdom (Grant Number EP/M024423/1) and the Carnegie Trust for the Universities of Scotland.
Finally, we gratefully acknowledge our colleagues whose shared interest in K$_2$CuF$_4$ enabled this work: Drs. Tetsuya Akashi, Hiroto Kasai, and Hiroyuki Shinada of Hitachi Ltd.; Prof. Yusuke Kousaka of Osaka Prefecture University; Prof. Jun-ichiro Kishine of The Open University of Japan; and Prof. Jun Akimitsu of Okayama University.

\renewcommand{\thefigure}{A\arabic{figure}}
\setcounter{figure}{0}  

\appendix
\section{Appendices}
\subsection{Automated defocus determination}
\begin{figure*}[thb]
  \centering
      \includegraphics[width=13cm]{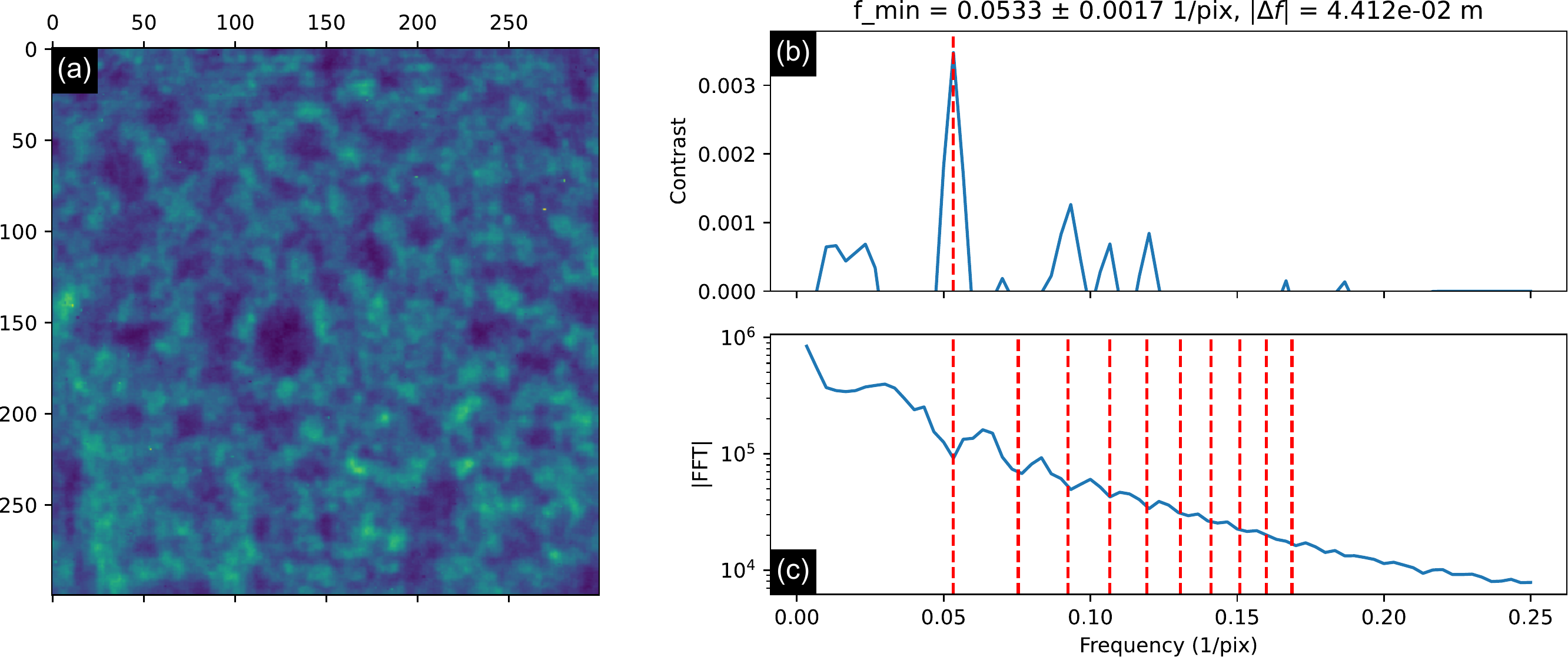}
      \caption{Example of automated defocus analysis.
      (a) Selected region of the image in Fig.~\ref{fig:TEM_images}(c), with axes labeled in pixels.
      (b) and (c) are results of processing (a) with the \texttt{fpd.tem\_tools.defocus\_from\_image} function.
      (b) A `contrast' parameter as a function of spatial frequency (a measure of defocus), with the red vertical dashed line indicating the automatically detected optimum value from the peak (see text for details).
      (c) Azimuthal average of the Fourier transform of (a).
      The red vertical dashed lines indicate the detected locations of the minima from the optimised defocus value.}
      \label{fig:defocus}
\end{figure*}

The defocus level of a Fresnel image is commonly estimated from the first zero crossing of the contrast transfer function (CTF), with the relevant spatial frequency determined from inspection of a Fourier transform of the spatially calibrated image.
In this work, we use multiple zeros of the CTF to obtain a generally more reliable and accurate estimate of the defocus in an automated procedure.
The procedure is implemented in the \texttt{defocus\_from\_image} function of the \texttt{tem\_tools} module of the open source \texttt{fpd} Python package~\citep{fpd}, and is described below with reference to Figure~\ref{fig:defocus}.

Figure~\ref{fig:defocus}(a) shows the region of the original Fresnel image selected for analysis, with the axes labeled in pixels.
This image is Fourier transformed and the result azimuthally averaged to produce the profile shown in Fig.~\ref{fig:defocus}(c).
Here, multiple minima can be seen; their location are indicated by the red vertical dashed lines.
To determine the optimum location of the minima, a contrast metric is calculated for all possible spatial frequencies of the first CTF minimum (a measure of the defocus value) within a specified range of the Nyquist frequency.
The metric value used is the mean of the contrast ($a-b / a+b$) at and between the predicted minima.
This metric is shown in Fig.~\ref{fig:defocus}(b), where a clear peak exists at the optimum location of the first CTF minimum.
The equivalent defocus value was determined automatically by finding the location of the peak, which is marked by a red vertical dashed line.
In this particular case, the defocus value was 44~mm, as indicated in the annotation of Fig.~\ref{fig:defocus}(b).

\subsection{Distortion Field Extraction}
In our previous analysis~\citep{Togawa2021_JPSJ_transition} of experimental K$_2$CuF$_4$ data using the theory outlined in Part~I of this work, we used the BUnwarpJ~\citep{bunwarpj_2006} plugin for the open source image processing package Fiji~\citep{fiji_2012}.
This plugin calculates elastic deformations between images on a basis represented by cubic B-splines, through iterative minimisation of a total energy term.
The term is primarily set by the dissimilarity between the source and target images, with optional regularisation terms based on the divergence and curl of the deformation fields (set to zero in our analysis).
While 16$\times$16 intervals in the grid of B-splines (the `super fine' setting of BUnwarpJ) adequately accounted for the deformations seen in the data, there was very limited control over the spatial resolution, and great care had to be taken around the edges of the sample.

The method and software tools for extraction of the distortion field reported in Part~I of this work overcomes the above limitations by using a different approach, that of non-rigid alignment of Fresnel images using a gradient descent method~\cite{Kroon2009_demon_reg}.
Advancement of this iterative algorithm was regulated by convolution of updates to the cumulative displacement field with a 2-D Gaussian distribution of standard deviation $\sigma$, which essentially controls the spatial resolution of the extracted data, up to any limits imposed by the microscope optics and sample related contributions to the same.
Custom regularisations may also be used in our analysis tools, but these tend to be slower than Gaussian regularisations and they also complicate interpretation of the spatial resolution.

The main scale of relevance for alignment to work correctly is the non-magnetic texture, which is comparable to the image distortion scale.
To reduce the probability of the alignment procedure getting stuck in local minima, the input images could have been initially filtered to remove higher frequency components.
Instead, the alignment was initially performed with a much larger $\sigma$ value, and the process repeated using the previous displacement field as the initial values for the next step, with $\sigma$ decreasing at each step (the first displacement field was taken as zero).
By doing so, we are in effect including only the lowest frequency components and then successively adding higher frequency ones.
In principle, a single alignment can be performed and the regularisation modified in real time according to the live feedback provided by the GUI.
However, the sequential approach described here allows the process to be automated and for us to observe the compromise between spatial resolution and the noise level by having access to the intermediate results.

\begin{figure*}[hbt!]
  \centering
      \includegraphics[width=16.5cm]{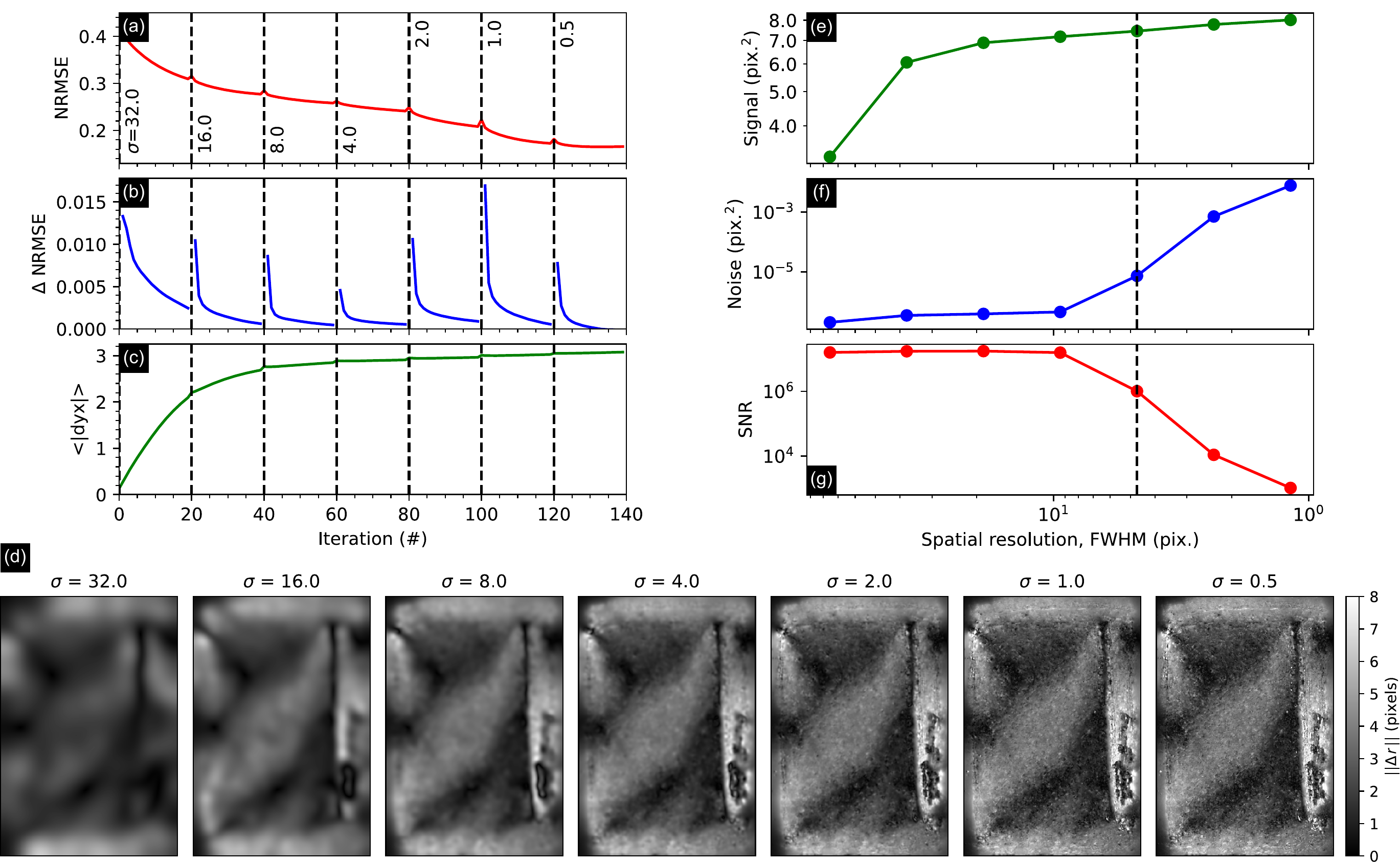}
      \caption{Non-rigid image alignment of ferromagnetic data at 5~K as the regularisation parameter $\sigma$ is reduced.
      A fixed number of 20 iterations was used for each $\sigma$ value.
      (a)-(c) Alignment metrics: (a) normalised root mean square error (NRMSE); (b) its change between iterations; and (c) the mean vector field magnitude.
      The small increases in (a) at the transitions are transients which result from stepped reductions in $\sigma$.
      (d) Displacement vector field magnitude maps for the same data.
      (e)-(g) Signal-to-noise ratio (SNR) analysis of the images in (d) with the support frame cropped.
      The values shown are per pixel and the data is plotted against spatial resolution.
      Note that the $x$-axis is reversed to match the data order shown in other panels.
      The data in the manuscript was produced using a final $\sigma$ value of 2 (marked by a vertical dashed line in (e)-(g)), which provides a good compromise between resolution and noise.}
      \label{fig:TEM_align}
\end{figure*}

Figure~\ref{fig:TEM_align} summarises the results of the alignment process as $\sigma$ is successively reduced by a factor of 2 from 32~pixels until it reaches 0.5~pixels.
The entire dataset was aligned simultaneously across multiple CPU cores, but for simplicity we show only the data at the base temperature of 5~K where the sample is ferromagnetic.
Figures~\ref{fig:TEM_align}(a) - \ref{fig:TEM_align}(c) display metrics during the alignment procedure.
Each vertical dashed line marks the end of a sequence of iterations at the constant $\sigma$ values shown in the annotations to (a).
To visualise the distortion field, we show maps of the displacement vector field magnitude at each of these points in the panels of Fig.~\ref{fig:TEM_align}(d).
The small area of discontinuous displacement field to the right of the sample arose from a local minima and went uncorrected.
Figures~\ref{fig:TEM_align}(e) - \ref{fig:TEM_align}(g) show the results of an analysis of the signal-to-noise ratio (SNR) in the images of Fig.~\ref{fig:TEM_align}(d), plotted against the spatial resolution.

As the alignment proceeds, the normalised root mean square error (NRMSE) [Fig.~\ref{fig:TEM_align}(a)] rapidly reduces and then tends towards a plateau at around $\sigma$ = 4~pixels, before decreasing again to a lower plateau.
At each step, the rate of change of the NRMSE [Fig.~\ref{fig:TEM_align}(b)] rapidly decreases and tends towards a plateau, indicating that faster alignment would have been achieved by adapting the number of iterations of the algorithm across the $\sigma$ values; we use a fixed number here for demonstration purposes.
In our particular case, the corner between the NRMSE plateaus occurs when the magnitude of the displacement field [Fig.~\ref{fig:TEM_align}(c)] has already reached close to its maximum value, and we thus interpret this feature as the point beyond which only small amplitude, high frequency signals are being added to the displacement field, and these are more likely to include noise or unwanted signal.
This interpretation is confirmed by visual inspection of the displacement maps [Fig.~\ref{fig:TEM_align}(d)] and the SNR analysis plots [Figs.~\ref{fig:TEM_align}(e) - \ref{fig:TEM_align}(g)], discussed below.

It can often be difficult to estimate the SNR ratio of single images with a high degree of confidence.
The signal power is simple to calculate and its value per pixel is shown in Fig.~\ref{fig:TEM_align}(e).
For the noise power, we adopt one of the methods we found success with previously~\citep{paterson_fpd2_2020}, based on wavelet shrinkage~\citep{Donoho_biomet_1994_noise_est} and provided by the scikit-image package~\citep{scikit-image}.
The noise power per pixel obtained from this analysis is shown in Fig.~\ref{fig:TEM_align}(f).
As expected from the images in the figure, the noise starts at a low level and then rapidly increases as $\sigma$ is reduced below 4 pixels.
We note that the extracted values will underestimate contributions to the apparent noise from small systematic imperfections in the alignment, which contributes to the slightly mottled appearance of the images.
Nevertheless, the extracted values still give an indication of the relative noise level in the data, and also demonstrate a useful method for characterising progression of the alignment.
The SNR values are shown in Fig.~\ref{fig:TEM_align}(g) and begin to decrease at a spatial resolution of $\sim$10 pixels.

For the properties of the sample magnetisation and images used here, $\sigma = 2$~pixels was used for the data presented in the manuscript. 
Each pixel was $\sim$10.46~nm in size, making the full-width-half-maximum resolution $\sim$49~nm.
This point is marked by the vertical dashed lines in Figs.~\ref{fig:TEM_align}(e) - \ref{fig:TEM_align}(g), and it  provides a reasonable compromise between spatial resolution and noise.
However, even smaller $\sigma$ values could be used and the data further processed to keep only the desired components.
These include edge-preserving filters such as a total variation (TV) regularisation, or ones which do not handle discontinuities well, such as smoothed splines, but only after cropping the data (TV or other custom filters may be applied during the alignment, but these tend to be slower than using a simple Gaussian filter).
That no cropping is required here is one of the significant advantages of our approach.
When dealing with multiple images, then any filtering may be improved by taking advantage of the additional dimension (temperature, in our case), and this also opens the possibility for the application of multivariate analyses such as dimensionality reduction.
The data in the manuscript was presented without any filtering.

\subsection{Phase Transition}
Figure~\ref{fig:temperature_dependence} shows TEM-DPC induction maps from the K$_2$CuF$_4$ sample at selected temperatures as it was heated through a phase transition with a 25~Oe field applied out-of-plane.
The supplemental video shows TEM-DPC induction maps at all temperatures between 5.0~K and 7.0~K, using an identical colour wheel for the vector field.

\normalsize

\bibliographystyle{MandM}

\balance

\end{document}